\documentclass[12pt,reqno]{amsart}
\usepackage{amsfonts}
\usepackage{graphicx}
\usepackage{amscd}
\usepackage{amsmath}
\usepackage{epsfig}
\usepackage[usenames]{color}
\usepackage{subfigure}
\usepackage[english]{babel}

\providecommand{\U}[1]{\protect\rule{.1in}{.1in}}
\providecommand{\U}[1]{\protect\rule{.1in}{.1in}}
\allowdisplaybreaks
\newtheorem{theorem}{Theorem}
\newtheorem{lemma}{Lemma}
\newtheorem{proposition}{Proposition}
\theoremstyle{plain}
\newtheorem{acknowledgement}{Acknowledgement}

\newtheorem{conclusion}{Conclusion}

\newtheorem{corollary}{Corollary}

\newtheorem{example}{Example}

\numberwithin{equation}{section}

\begin{document}
\title[ Schr$\ddot{\mbox{O}}$dinger Equation with Quadratic Hamiltonian]{On
persistence of superoscillations for the Schr\"{o}dinger equation with
time-dependent quadratic Hamiltonians}
\author{E. Hight}
\author{T. Oraby}
\author{J. Palacio}
\author{E. Suazo}
\address{School of Mathematical and Statistical Sciences, University of
Texas Rio Grande Valley, 1201 W. University Drive, Edinburg, Texas,
78539-2999.}
\email{erwin.suazo@utrgv.edu}
\email{jose.palacio01@utrgv.edu}
\email{tamer.oraby@utrgv.edu}
\email{elijah.hight01@utrgv.edu}
\date{\today }

\begin{abstract}
In this work we study the persistence in time of superoscillations for the
Schr\"{o}dinger equation with quadratic time-dependent Hamiltonians. We have
solved explicitly the Cauchy initial value problem with three different kind
of oscillatory initial data. In order to prove the persistence of
superoscillations we have defined explicitly an operator in terms of
solutions of a Riccati system associated with the variable coefficients of
the Hamiltonian. The operator is defined on a space of entire functions.
Particular examples include Caldirola-Kanai and degenerate parametric
harmonic oscillator Hamiltonians and more. For these examples we have
illustrated numerically the convergence on real and imaginary parts.\newline
\end{abstract}

\maketitle


\section{ Introduction}

Since 1964 when the work of Aharonov\footnote{%
Aharonov is also well-known by the Aharonov-Bohm effect.} and collaborators 
\cite{Aha0}, \cite{Aha01} appeared, quantum physicists were attracted to and
experimentally demonstrated the superoscillations phenomena; for an
excellent review see \cite{Aha3}, also see \cite{Aha0}-\cite{Aha5}. Aharonov
et al. have shown that superoscillations naturally arise when dealing with
weak values, providing a fundamentally different way to make measurements in
quantum physics. Superoscillating functions have attracted the attention of
mathematicians by the superposition of small Fourier components with a
bounded Fourier spectrum. Applications include antenna theory, metrology and
a new theory of superresolution in optics, see the work of Berry \cite%
{Berry1}- \cite{Berry7}, Lindberg \cite{Lindberg} and references therein.

A natural question arises: What is the most general time dependent
Hamiltonian $H(t)$ for which the Schr\"{o}dinger equation presents
persistence in time of superoscillations? In other words, if we define $\psi
_{n}(x,t)$ as the solution of 
\begin{eqnarray*}
{\tiny i}\frac{\partial \psi _{n}(x,t)}{\partial t} &=&H(t)\psi _{n}(x,t) \\
\psi _{n}(x,0) &=&\left( \cos (x/n)+ia\sin (x/n)\right) ^{n}
\end{eqnarray*}

and define $\psi _{a}(x,t)$ as the solution of 
\begin{eqnarray*}
{\tiny i}\frac{\partial \psi _{a}(x,t)}{\partial t} &=&H(t)\psi _{a}(x,t) \\
\psi _{a}(x,0) &=&e^{iax},
\end{eqnarray*}%
since $\lim_{n\rightarrow \infty }\left( \cos (x/n)+ia\sin (x/n)\right)
^{n}=e^{iax},$ (See Figure 1), do we have $\lim_{n\rightarrow \infty }\psi
_{n}(x,t)=\psi _{a}(x,t)?$

In this work we prove that the generalized harmonic oscillator of the form%
\begin{equation}
i\partial _{t}\psi =-a\left( t\right) \partial _{x}^{2}\psi +b\left(
t\right) x^{2}\psi -ic\left( t\right) x\partial _{x}\psi -id\left( t\right)
\psi -f(t)x\psi +ig(t)\partial _{x}\psi  \label{SE1}
\end{equation}%
does satisify this property, and further we provide examples with explicit
constructions of the Green functions. We also provide numerical simulations
for better understanding of this phenomena. The generalized harmonic
oscillator (\ref{SE1}) has attracted considerable attention over many years
in view of its great importance to several advanced quantum problems,
including Berry's phase, quantization of mechanical of systems and more (see 
\cite{Co:Suslov} and references therein). The fact that in quantum
electrodynamics the electromagnetic field can be represented as a set of
forced harmonic oscillators makes quadratic Hamiltonians of special interest 
\cite{DoMaMa, Hannay, YeLeePan}. A method to construct explicit propagators
for the linear Schr\"{o}dinger equation with a time-dependent quadratic
Hamiltonian based in solutions of the Riccati equation has been presented in 
\cite{Cor-Sot:Lop:Sua:Sus}.

The quantum harmonic oscillator is probably the most beautiful example to
introduce the theory of superoscillations, see \cite{Aha3}, mainly because
of the convenient use of Mehler's formula for the Green function (or Feynman
propagator \cite{Fey:Hib}). In \cite{Cor-Sot:Lop:Sua:Sus}, \cite{Lo:Su:VeSy}-%
\cite{Sua1} Suslov and collaborators introduced a generalization of Mehler's
formula. The main result of this work is to prove superoscillations where
the generalized Mehler's formula \cite{Cor-Sot:Lop:Sua:Sus} can be applied
for certain kinds of variable quadratic Hamiltonians. Therefore, in the
present work we study the superoscillations for the Schr\"{o}dinger equation
with variable coefficients of the form (\ref{SE1}).

We have performed numerical calculations of limits and numerical simulations
of solutions of some Schr\"{o}dinger equations. For the limits problems, we
used the point-wise difference between the real as well as imaginary parts
of a sequence of functions and their limiting functions. We showed that both
of the differences go to zero. We solved also a Schr\"{o}dinger equation
using finite difference method over space along with Runge-Kutta method over
time. We are showing visually how there are in agreement.

This paper is organized as follows: In Section 2 we review explicit
solutions for the Riccati system (\ref{rica1})-(\ref{rica6}) that we will
use to solve the Cauchy initial value problem for the Schr\"{o}dinger
equation (\ref{SE1}) with oscillatory initial data. For this purpose we also
review the general form of the Green function for (\ref{SE1}). At the end of
this section we review a fundamental theorem on the convergence of
convolution of operators on a space of entire functions. In Section 3 we
prove the most important result of this work: the superoscillations for the
Schr\"{o}dinger equation with variable coefficients of the form (\ref{SE1})
persist on time. In Section 4, we present several relevant examples
including Caldirola-Kanai, modified Caldirola-Kanai, degenerate parametric
harmonic oscillator and Meiler, Cordero-Soto, Suslov Hamiltonians. Finally,
we have added an appendix explaining the solution of a Ince's type equation,
relevant for the example of a degenerate parametric oscillator.

\section{Preliminary results}

Our main result will need explicit solutions for the Riccati system (\ref%
{rica1})-(\ref{rica6}). Therefore, we need the following Lemma:

\begin{lemma}
\cite{Cor-Sot:Lop:Sua:Sus}, \cite{Co:Suslov} Assuming that $a(t),b(t)$, $%
c(t) $, $d(t),$ $f(t)$ and $g(t)$ are piecewise real continuous functions,
there exists an interval $I$ of time where the following (Riccati-type)
system \ 
\begin{equation}
\dfrac{d\alpha }{dt}+b(t)+2c(t)\alpha +4a(t)\alpha ^{2}=0,  \label{rica1}
\end{equation}%
\begin{equation}
\dfrac{d\beta }{dt}+(c(t)+4a(t)\alpha (t))\beta =0,  \label{rica2}
\end{equation}%
\begin{equation}
\dfrac{d\gamma }{dt}+a(t)\beta ^{2}(t)=0,  \label{rica3}
\end{equation}%
\begin{equation}
\dfrac{d\delta }{dt}+(c(t)+4a(t)\alpha (t))\delta =f(t)+2\alpha (t)g(t),
\label{rica4}
\end{equation}%
\begin{equation}
\dfrac{d\varepsilon }{dt}=(g(t)-2a(t)\delta (t))\beta (t),  \label{rica5}
\end{equation}%
\begin{equation}
\dfrac{d\kappa }{dt}=g(t)\delta (t)-a(t)\delta ^{2}(t)  \label{rica6}
\end{equation}%
\ 

has an explicit solution given by\ 
\begin{equation}
\alpha \left( t\right) =\frac{1}{4a\left( t\right) }\frac{\mu _{0}^{\prime
}\left( t\right) }{\mu _{0}\left( t\right) }-\frac{d\left( t\right) }{%
2a\left( t\right) },  \label{alpha0}
\end{equation}%
\begin{equation}
\beta \left( t\right) =-\frac{w\left( t\right) }{\mu _{0}\left( t\right) }%
,\quad w\left( t\right) =\exp \left( -\int_{0}^{t}\left( c\left( s\right)
-2d\left( s\right) \right) \ ds\right) ,  \label{beta0}
\end{equation}%
\begin{equation}
\gamma \left( t\right) =\frac{d\left( 0\right) }{2a\left( 0\right) }+\frac{1%
}{2\mu _{1}\left( 0\right) }\frac{\mu _{1}\left( t\right) }{\mu _{0}\left(
t\right) },  \label{gamma0}
\end{equation}%
\begin{equation}
\delta \left( t\right) =\frac{w\left( t\right) }{\mu _{0}\left( t\right) }\
\ \int_{0}^{t}\left[ \left( f\left( s\right) -\frac{d\left( s\right) }{%
a\left( s\right) }g\left( s\right) \right) \mu _{0}\left( s\right) +\frac{%
g\left( s\right) }{2a\left( s\right) }\mu _{0}^{\prime }\left( s\right) %
\right] \ \frac{ds}{w\left( s\right) },  \label{delta0}
\end{equation}%
\begin{eqnarray}
\varepsilon \left( t\right) &=&-\frac{2a\left( t\right) w\left( t\right) }{%
\mu _{0}^{\prime }\left( t\right) }\delta _{0}\left( t\right) +8\int_{0}^{t}%
\frac{a\left( s\right) \sigma \left( s\right) w\left( s\right) }{\left( \mu
_{0}^{\prime }\left( s\right) \right) ^{2}}\left( \mu _{0}\left( s\right)
\delta _{0}\left( s\right) \right) \ ds  \label{epsilon0} \\
&&+2\int_{0}^{t}\frac{a\left( s\right) w\left( s\right) }{\mu _{0}^{\prime
}\left( s\right) }\left[ f\left( s\right) -\frac{d\left( s\right) }{a\left(
s\right) }g\left( s\right) \right] \ ds,  \notag
\end{eqnarray}%
\begin{eqnarray}
\kappa \left( t\right) &=&\frac{a\left( t\right) \mu _{0}\left( t\right) }{%
\mu _{0}^{\prime }\left( t\right) }\delta _{0}^{2}\left( t\right)
-4\int_{0}^{t}\frac{a\left( s\right) \sigma \left( s\right) }{\left( \mu
_{0}^{\prime }\left( s\right) \right) ^{2}}\left( \mu _{0}\left( s\right)
\delta _{0}\left( s\right) \right) ^{2}\ ds  \label{kappa0} \\
&&\quad -2\int_{0}^{t}\frac{a\left( s\right) }{\mu _{0}^{\prime }\left(
s\right) }\left( \mu _{0}\left( s\right) \delta _{0}\left( s\right) \right) %
\left[ f\left( s\right) -\frac{d\left( s\right) }{a\left( s\right) }g\left(
s\right) \right] \ ds,  \notag
\end{eqnarray}%
\ with $\delta \left( 0\right) =g\left( 0\right) /\left( 2a\left( 0\right)
\right) ,$ $\varepsilon \left( 0\right) =-\delta \left( 0\right) ,$ $\kappa
\left( 0\right) =0.$ Here $\mu _{0}$ and $\mu _{1}$ represent the
fundamental solution of the characteristic equation \ 
\begin{equation}
\mu ^{\prime \prime }-\tau (t)\mu ^{\prime }+4\sigma (t)\mu =0,
\label{CharacteristicEquatrion1}
\end{equation}%
with\ 
\begin{equation}
\tau (t)=\frac{a^{\prime }}{a}-2c+4d,\hspace{1cm}\sigma (t)=ab-cd+d^{2}+%
\frac{d}{2}\left( \frac{a^{\prime }}{a}-\frac{d^{\prime }}{d}\right)
\label{CharacteristicEquatrion2}
\end{equation}

subject to the initial conditions $\mu _{0}(0)=0$, $\mu _{0}^{\prime
}(0)=2a(0)\neq 0$ and $\mu _{1}(0)\neq 0$, $\mu _{1}^{\prime }(0)=0$.
\end{lemma}

Also, we will need the following Theorem to solve the Cauchy initial value
problem with oscillatory initial data.

\begin{theorem}
\cite{Cor-Sot:Lop:Sua:Sus} The Green function, or Feynman's propagator,
corresponding to the Schr\"{o}dinger equation (\ref{SE1}) can be obtained as 
\begin{equation}
\psi =G\left( x,y,t\right) =\frac{1}{\sqrt{2\pi i\mu \left( t\right) }}\
e^{i\left( \alpha \left( t\right) x^{2}+\beta \left( t\right) xy+\gamma
\left( t\right) y^{2}+\delta (t)x+\varepsilon \left( t\right) y+\kappa
(t)\right) },  \label{newgreen}
\end{equation}%
where $\alpha \left( t\right) ,$ $\beta \left( t\right) ,$ $\gamma \left(
t\right) ,$ $\delta (t),$ $\varepsilon \left( t\right) $ and $\kappa (t)$\
are solutions of the Riccati-type system (\ref{rica1})-(\ref{rica6}). Then
the superposition principle allows us to solve the corresponding Cauchy
initial value problem; the solution is given by%
\begin{equation*}
\psi (x,t)=\int_{-\infty }^{\infty }G(x,y,t)\psi (y,0)dy
\end{equation*}%
for suitable data $\psi (x,0)=\varphi (x).$
\end{theorem}

In order to prove the persistence of superoscillations we will define a
convolution operator in the following space of entire functions. For the
proof of the following Lemma, see \cite{Aha3}.

\begin{lemma}
Let's consider the class $A_{1}$ as the set of entire functions such that
there exists $A>0$ and $B>0$ for which%
\begin{equation}
\left\vert f(z)\right\vert \leq Ae^{B\left\vert z\right\vert }
\end{equation}%
for all $z\in 
\mathbb{C}
.$ Let $\lambda (t)$ be a complex valued bounded function for $t\in \lbrack
0,T]$ for some $T\in (0,\infty )$ and let $f\in A_{1}.$ Then, for $p\in 
\mathbb{N}
$%
\begin{equation}
P_{\lambda }(t,\partial _{z})f=\sum_{n=0}^{\infty }\frac{\lambda (t)^{n}}{n!}%
\partial _{z}^{pn}f\in A_{1}.
\end{equation}%
Further, $P_{\lambda }(t,\partial _{z})$ is continous on $A_{1},$ that is, $%
P_{\lambda }(t,\partial _{z})f\rightarrow 0$ as f$\rightarrow 0.$
\end{lemma}

\textbf{Assumption 1. }$a(t),b(t)$, $c(t)$, $d(t),$ $f(t)$ and $g(t)$ are
suitable functions such that%
\begin{equation}
\rho (t)=\frac{i\gamma (h^{2}-\varepsilon ^{2}-4\kappa \gamma h-2\varepsilon
h)}{(2\delta \gamma +\beta h+\beta \varepsilon )^{2}}
\end{equation}

is a complex valued bounded function for $t\in \lbrack 0,T]$ for some $T\in
(0,\infty ).$ And $\beta \left( t\right) ,$ $\gamma \left( t\right) ,$ $%
\delta (t),$ $\varepsilon \left( t\right) $ and $\kappa (t)$\ are solutions
of the Riccati-type system (\ref{rica2})-(\ref{rica6}) given by (\ref{beta0}%
)-(\ref{kappa0}).

\section{Persistence of Superoscillations for the Schr\"{o}dinger equation
with variable coefficients}

The following is our main result:

\begin{theorem}
If the characteristic equation (\ref{CharacteristicEquatrion1})-(\ref%
{CharacteristicEquatrion2}) associated to the variable coefficient Schr\"{o}%
dinger equation (\ref{SE1}) admits two standard solutions $\mu _{0}$ and $%
\mu _{1}$ subject to%
\begin{equation}
\mu _{0}\left( 0\right) =0,\quad \mu _{0}^{\prime }\left( 0\right) =2a\left(
0\right) \neq 0\qquad \mu _{1}\left( 0\right) \neq 0,\quad \mu _{1}^{\prime
}\left( 0\right) =0,
\end{equation}

then

1. The solution for the Cauchy initial value problem for (\ref{SE1}) subject
to $\psi (x,0)=e^{ihx}$ is given by%
\begin{equation}
\phi _{h}(x,t)=\frac{1}{\sqrt{2\mu \gamma }}e^{i\left( 4\alpha \gamma -\beta
^{2}\right) x^{2}/4\gamma }e^{i\left( \delta \gamma /h+\beta /2-\beta
\varepsilon /2h\right) hx/\gamma }e^{-i\left( h^{2}-\varepsilon ^{2}-4\kappa
\gamma -2\varepsilon h\right) /4\gamma },
\end{equation}

where $\alpha \left( t\right) ,$ $\beta \left( t\right) ,$ $\gamma \left(
t\right) ,$ $\delta (t),$ $\varepsilon \left( t\right) $ and $\kappa (t)$\
are solutions of the Riccati-type system (\ref{rica1})-(\ref{rica6}).

2. The solution for the Cauchy initial value problem for (\ref{SE1}) subject
to 
\begin{equation}
\psi (x,0)=F_{n}(x,h)=\left( \cos \left( \frac{x}{n}\right) +ih\sin \left( 
\frac{x}{n}\right) \right) ^{n}=\sum_{k=0}^{n}C_{k}(n,h)e^{ix(1-2k/n)}
\end{equation}

is given by%
\begin{equation}
\psi _{n}(x,t)=\sum_{k=0}^{n}C_{k}(n,h)\phi _{1-\frac{2k}{n}}(x,t),
\end{equation}

where $C_{k}(n,h)=\binom{n}{k}\left( \frac{1+h}{2}\right) ^{n-k}\left( \frac{%
1-h}{2}\right) ^{k}$.\newline
\begin{figure}[h]
\begin{centering}
{\includegraphics[scale=0.25]{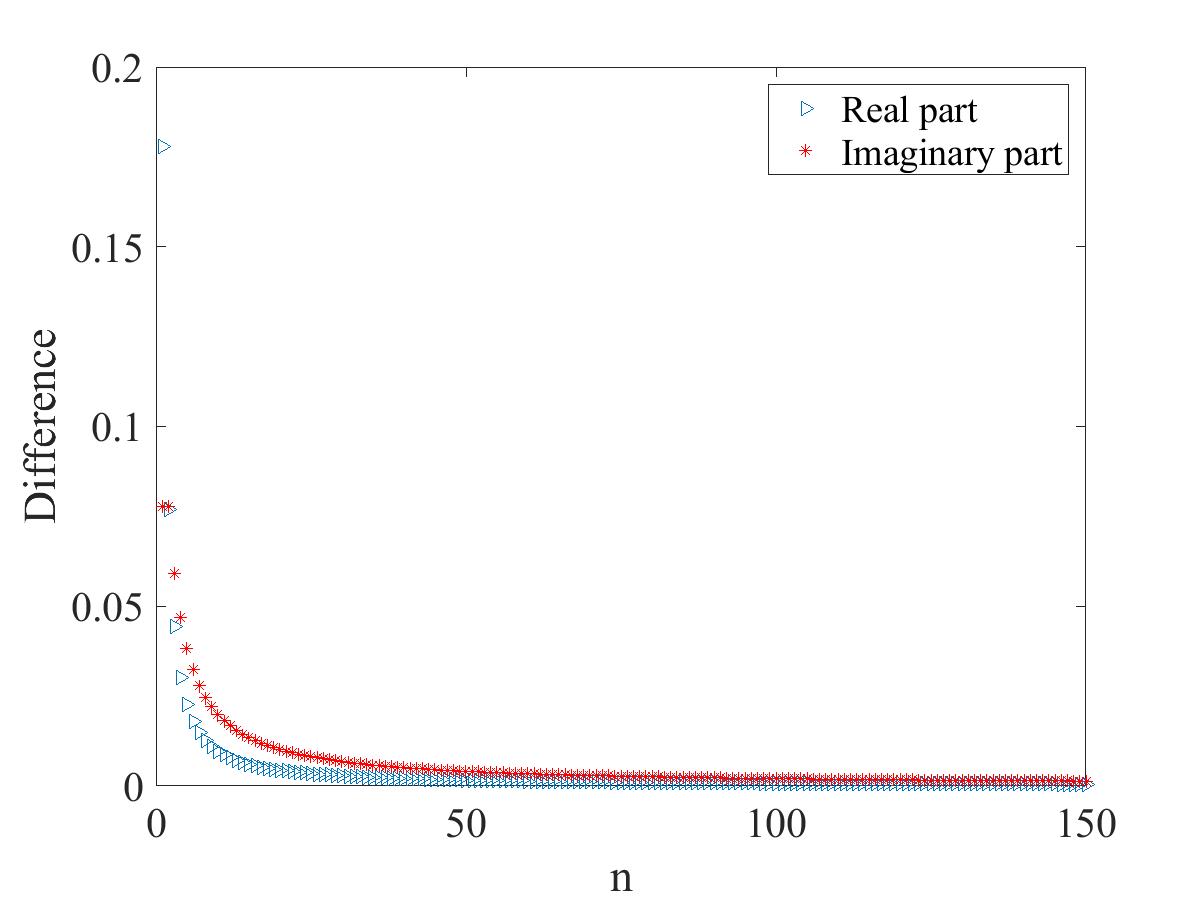}}
\par\end{centering}
\caption{Limit of the difference between the real parts (in blue color) and
imaginary parts (in red color) of $F_{n}(x,h)=\left( \cos \left( \frac{x}{n}%
\right) +ih\sin \left( \frac{x}{n}\right) \right) ^{n}$ and $e^{ihx}$ at
different values of $n$. It was calculated for $h=1.2$ and $x=1$.}
\label{fig1}
\end{figure}
3. If the coefficients of (\ref{SE1}) satisfy Assumption 1, the
superoscillations for (\ref{SE1}) persist on time, i.e.%
\begin{equation}
\lim_{n\rightarrow \infty }\psi _{n}(x,t)=\phi _{h}(x,t),h>1.  \label{limit}
\end{equation}
\end{theorem}

In order to prove Theorem 2, we need to prove the following Lemma first:

\begin{lemma}
The solution of $\phi _{h}(x,t)$ can be represented as%
\begin{equation}
\phi _{h}(x,t)=\frac{1}{\sqrt{2\mu _{0}\gamma }}e^{i\left( \frac{4\alpha
\gamma -\beta ^{2}}{4\gamma }\right) x^{2}}U\left( t,\frac{d}{dx}\right) %
\left[ e^{i\left( \delta +\beta \varepsilon /2\gamma +h\beta /\gamma \right)
x}\right] ,
\end{equation}

where we define%
\begin{equation}
U\left( t,\frac{d}{dx}\right) =\sum_{m\geq 0}\frac{1}{m!}\left[ \frac{%
i\gamma (h^{2}-\varepsilon ^{2}-4k\gamma h-2\varepsilon h)}{(2\delta \gamma
+\beta h+\beta \varepsilon )^{2}}\right] ^{m}\frac{d^{2m}}{dx^{2m}}.
\label{ConvOpe}
\end{equation}
\end{lemma}

\textbf{Proof:} By definition, by Lemma 1 and Theorem 1 we have 
\begin{equation*}
\phi _{h}(x,t)=\int_{\mathbb{R}}G(x,y,t)e^{ihy}dy=\int_{\mathbb{R}}\frac{1}{%
\sqrt{2\pi i\mu _{0}}}e^{i(\alpha (t)x^{2}+\beta (t)xy+\gamma
(t)y^{2}+\delta (t)x+\varepsilon (t)y+\kappa (t))}e^{ihy}dy.
\end{equation*}%
By the explicit expressions (\ref{alpha0})-(\ref{kappa0}), by the standard
formula%
\begin{equation*}
\int_{\mathbb{R}}e^{i\left[ Ay^{2}+2By\right] }dy=\sqrt{\frac{i\pi }{A}}e^{-i%
\frac{B^{2}}{A}},\quad Im(A)\leq 0,
\end{equation*}%
and by using power series expansion we get%
\begin{eqnarray*}
\phi _{h}(x,t) &=&\frac{1}{\sqrt{2\mu \gamma }}e^{i\left( 4\alpha \gamma
-\beta ^{2}\right) x^{2}/4\gamma }e^{i\left( \delta \gamma /h+\beta /2-\beta
\varepsilon /2h\right) hx/\gamma }e^{-i\left( h^{2}-\varepsilon ^{2}-4\kappa
\gamma -2\varepsilon h\right) /4\gamma } \\
&=&\frac{1}{\sqrt{2\mu \gamma }}e^{i\left( 4\alpha \gamma -\beta ^{2}\right)
x^{2}/4\gamma }\sum_{m\geq 0}\frac{1}{m!}\left( -i\frac{(1+\varepsilon
^{2}/h^{2}-4\kappa \gamma /h^{2}+2\varepsilon /h)}{4\gamma }\right)
^{m}h^{2m}e^{i\left( \delta \gamma /h-\beta /2-\beta \varepsilon /2h\right)
hx/\gamma }.
\end{eqnarray*}

To prove our Theorem 2 we also need the following proposition which can be
proved by induction.

\begin{proposition}
The following equality holds for $m\geq 0:$%
\begin{equation}
h^{2m}e^{i\left( \delta \gamma /h-\beta /2-\beta \varepsilon /2h\right)
hx/\gamma }=\left[ \frac{\gamma }{i\left( \delta \gamma /h-\beta /2-\beta
\varepsilon /2h\right) }\right] ^{2m}\frac{d^{2m}}{dx^{2m}}e^{i\left( \delta
\gamma /h-\beta /2-\beta \varepsilon /2h\right) hx/\gamma }.
\end{equation}
\end{proposition}

Therefore, by proposition 1 we obtain%
\begin{eqnarray}
\phi _{h}(x,t) &=&\frac{1}{\sqrt{2\mu \gamma }}e^{i\left( 4\alpha \gamma
-\beta ^{2}\right) x^{2}/4\gamma }\times \\
&&\sum_{m\geq 0}\frac{1}{m!}\left( i\frac{\gamma (1+\varepsilon
^{2}/h^{2}-4\kappa \gamma /h^{2}+2\varepsilon /h)}{4\left( \delta \gamma
/h-\beta /2-\beta \varepsilon /2h\right) ^{2}}\right) ^{m}\frac{d^{2m}}{%
dx^{2m}}e^{i\left( \delta \gamma /h-\beta /2-\beta \varepsilon /2h\right)
hx/\gamma }  \notag \\
&=&\frac{1}{\sqrt{2\mu \gamma }}e^{i\left( 4\alpha \gamma -\beta ^{2}\right)
x^{2}/4\gamma }U\left( t,\frac{d}{dx}\right) e^{i\left( \delta -h\beta
/2\gamma -\beta \varepsilon /2\gamma \right) x}.  \label{phia}
\end{eqnarray}

\textbf{Proof of the Theorem 2}: By (\ref{phia}), by Lemma 1, Theorem 1 and
by Lemma 2, we obtain 
\begin{align*}
\lim_{n\rightarrow \infty }\psi _{n}(x,t)& =\lim_{n\rightarrow \infty
}\sum_{k=0}^{n}C_{k}(n,h)\phi _{1-\frac{2k}{n}}(x,t) \\
& =\frac{1}{\sqrt{2\mu \gamma }}e^{i\left( 4\alpha \gamma -\beta ^{2}\right)
x^{2}/4\gamma }\lim_{n\rightarrow \infty }\sum_{k=0}^{n}C_{k}(n,h)U\left( t,%
\frac{d}{dx}\right) e^{i\delta x}e^{-i(1-2k/n)\beta x/\gamma }e^{-i\beta
\varepsilon x/2\gamma } \\
& =\frac{1}{\sqrt{2\mu \gamma }}e^{i\left( 4\alpha \gamma -\beta ^{2}\right)
x^{2}/4\gamma }\lim_{n\rightarrow \infty }U\left( t,\frac{d}{dx}\right)
e^{i\delta x}e^{-i\beta \varepsilon x/2\gamma }F_{n}\left( \frac{-x}{2\gamma
/\beta }\right) \\
& =\frac{1}{\sqrt{2\mu \gamma }}e^{i\left( 4\alpha \gamma -\beta ^{2}\right)
x^{2}/4\gamma }U\left( t,\frac{d}{dx}\right) e^{i(\delta -\beta \varepsilon
/2\gamma -h\beta /2\gamma )x} \\
& =\phi _{h}(x,t).
\end{align*}

\begin{corollary}
If the characteristic equation (\ref{CharacteristicEquatrion1})-(\ref%
{CharacteristicEquatrion2}) associated to the variable coefficient Schr\"{o}%
dinger equation 
\begin{equation}
i\partial _{t}\psi =-a\left( t\right) \partial _{x}^{2}\psi +b\left(
t\right) x^{2}\psi -ic\left( t\right) x\partial _{x}\psi -id\left( t\right)
\psi  \label{SE2}
\end{equation}%
admits two standard solutions $\mu _{0}$ and $\mu _{1}$ subject to%
\begin{equation}
\mu _{0}\left( 0\right) =0,\quad \mu _{0}^{\prime }\left( 0\right) =2a\left(
0\right) \neq 0\qquad \mu _{1}\left( 0\right) \neq 0,\quad \mu _{1}^{\prime
}\left( 0\right) =0,  \label{initial conditions}
\end{equation}%
the superoscillations for (\ref{SE2}) persist on time$.$
\end{corollary}

\textbf{Proof:} By Lemma 2, the convolution operator (\ref{ConvOpe}) becomes%
\begin{equation}
U\left( t,\frac{d}{dx}\right) =\sum_{m\geq 0}\frac{1}{m!}\left( i\gamma
(t)\mu _{0}^{2}\left( t\right) w^{-2}(t)\right) ^{m}\frac{d^{2m}}{dx^{2m}}%
\left[ e^{i\frac{(1-2k/n)x}{\mu _{1}}}\right] ,
\end{equation}

and also%
\begin{eqnarray*}
\phi _{h}(x,t) &=&\frac{1}{\sqrt{2\mu _{1}}}e^{i\left[ \left( \frac{\mu
_{0}^{^{\prime }}(t)}{\mu _{0}(t)}-\frac{1}{\mu _{0}\mu _{1}}\right) \right]
x^{2}}e^{i\left( \frac{hx}{\mu _{1}}-\frac{h^{2}\mu _{0}}{4\mu _{1}}\right) }
\\
&=&\frac{1}{\sqrt{2\mu _{1}}}e^{i\left[ \left( \frac{\mu _{0}^{^{\prime }}(t)%
}{\mu _{0}(t)}-\frac{1}{\mu _{0}\mu _{1}}\right) \right] x^{2}}U\left( t,%
\frac{d}{dx}\right) \left[ e^{i\frac{hx}{\mu _{1}}}\right] .
\end{eqnarray*}

Further, by Lemma 2 
\begin{eqnarray*}
\lim_{n\rightarrow \infty }\psi _{n}(x,t) &=&\lim_{n\rightarrow \infty
}\sum_{k=0}^{n}C_{k}(n,h)\frac{1}{\sqrt{2\mu _{1}}}e^{i\left[ \left( \frac{%
\mu _{0}^{^{\prime }}(t)}{\mu _{0}(t)}-\frac{1}{\mu _{0}\mu _{1}}\right) %
\right] x^{2}}e^{i\left( \frac{(1-\frac{2k}{n})x}{\mu _{1}}-\frac{(1-\frac{2k%
}{n})^{2}\mu _{0}}{4\mu _{1}}\right) } \\
&=&\frac{1}{\sqrt{2\mu _{1}}}e^{i\left[ \left( \frac{\mu _{0}^{^{\prime }}(t)%
}{\mu _{0}(t)}-\frac{1}{\mu _{0}\mu _{1}}\right) \right] x^{2}}U\left( t,%
\frac{d}{dx}\right) \lim_{n\rightarrow \infty }\sum_{k=0}^{n}C_{k}(n,h)\left[
e^{i\frac{(1-2k/n)x}{\mu _{1}}}\right] \\
&=&\frac{1}{\sqrt{2\mu _{1}}}e^{i\left[ \left( \frac{\mu _{0}^{^{\prime }}(t)%
}{\mu _{0}(t)}-\frac{1}{\mu _{0}\mu _{1}}\right) \right] x^{2}}U\left( t,%
\frac{d}{dx}\right) e^{ihx/\mu _{1}} \\
&=&\phi _{h}(x,t).
\end{eqnarray*}

\section{Some special cases}

In this section we apply the results of the previous section to several
models of the quantum damped oscillators in a framework of a general
approach to the time-dependent Schrodinger equation with variable quadratic
Hamiltonians, see \cite{Co:Su:Su}.

For further illustration, we will verify numerically this convergence: If $%
\phi _{h}(x,t)$ is given by%
\begin{eqnarray}
\phi _{h}(x,t) &=&\frac{1}{\sqrt{2\mu _{0}\gamma }}e^{i\left[ \left( 4\alpha
\gamma -\beta ^{2}\right) x^{2}/4\gamma +\beta hx/2\gamma -h^{2}/4\gamma %
\right] } \\
&=&\frac{1}{\sqrt{2\mu _{0}\gamma }}\cos \left( \left( 4\alpha \gamma -\beta
^{2}\right) x^{2}/4\gamma +\beta hx/2\gamma -h^{2}/4\gamma \right) \\
&&+i\frac{1}{\sqrt{2\mu _{0}\gamma }}\sin \left( \left( 4\alpha \gamma
-\beta ^{2}\right) x^{2}/4\gamma +\beta hx/2\gamma -h^{2}/4\gamma \right) . 
\notag
\end{eqnarray}%
we must have%
\begin{equation}
\lim_{n\rightarrow 0}\psi _{n}(x,t)=\lim_{n\rightarrow
0}\sum_{k=0}^{n}C_{k}(n,h)\phi _{1-\frac{2k}{n}}(x,t)=\phi _{h}(x,t)
\end{equation}%
where $C_{k}(n,h)=\binom{n}{k}\left( \frac{1+h}{2}\right) ^{n-k}\left( \frac{%
1-h}{2}\right) ^{k}.$ Indeed, that is shown numerically in the figures of
this section.

\begin{figure}[h]
\begin{centering}
\subfigure[Example 1]{\includegraphics[scale=0.2]{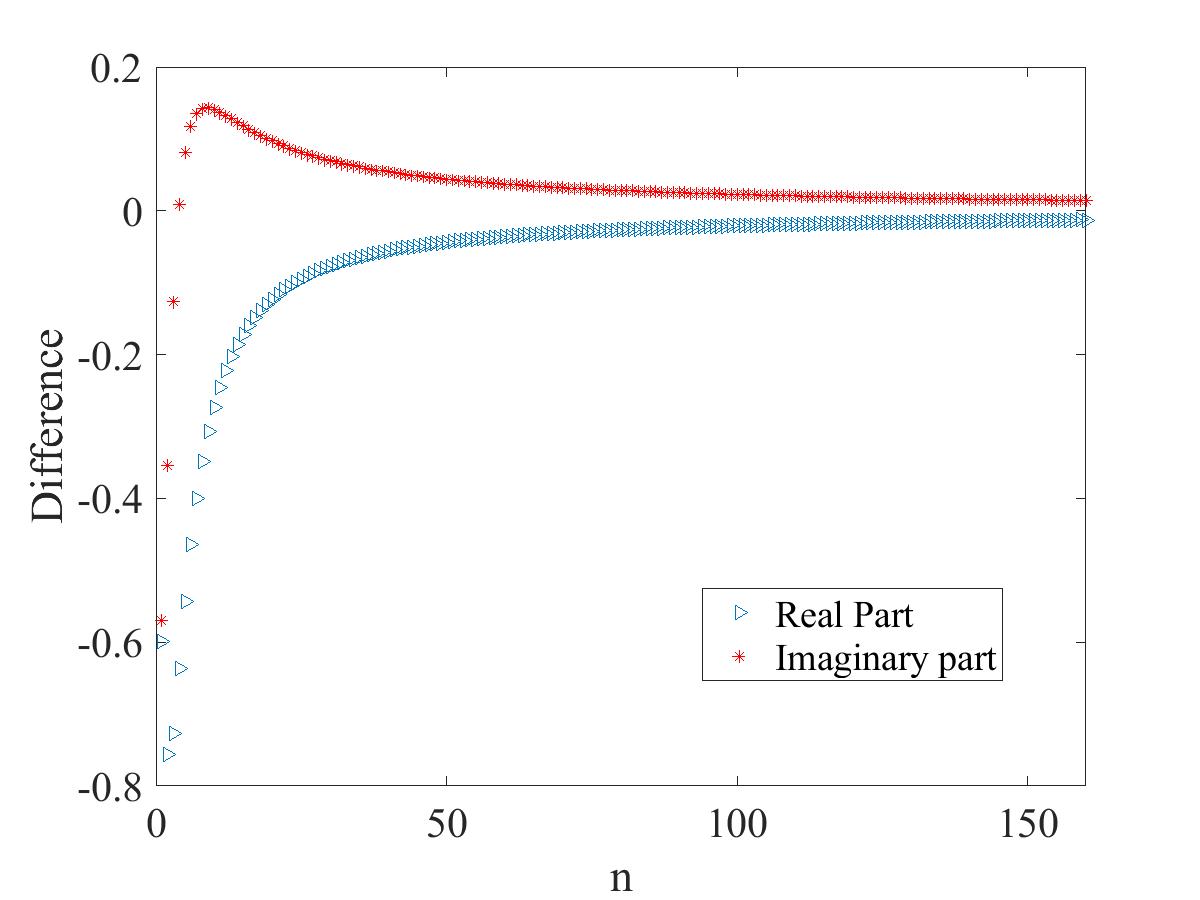}}\subfigure[Example 3]{\includegraphics[scale=0.2]{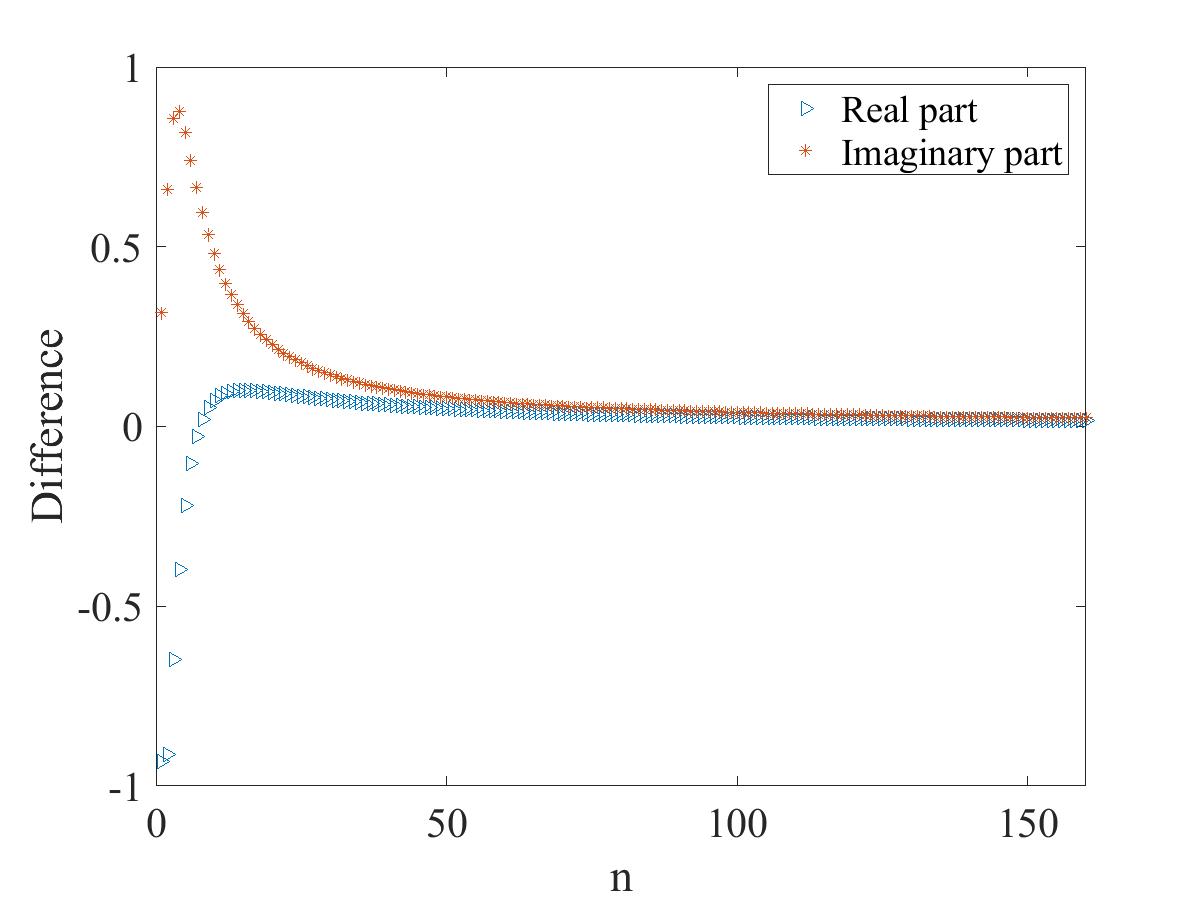}}
\par\end{centering}
\caption{Limit of the difference between the real parts (in blue color) and
imaginary parts (in red color) of the solution at different values of $n$.
It was calculated for (a) Example 1 and (b) $\protect\lambda =.1$ in Example
3, $h=1.2$, $x=1$, and $t=1$.}
\label{fig2}
\end{figure}

The first example illustrating superoscillations is of course the quantum
harmonic oscillator:

\begin{example}
The quantum harmonic oscillator 
\begin{equation*}
i\frac{\partial \psi }{\partial t}+\frac{1}{2}\frac{\partial ^{2}\psi }{%
\partial x^{2}}-x^{2}\psi =0
\end{equation*}

and its Green function (Mehler's formula) is given by 
\begin{equation}
G(x,y,t)=\frac{1}{\sqrt{2\pi i\sin t}}\exp \left( i\alpha (t)x^{2}+\beta
(t)xy+\gamma (t)y^{2}\right) ,\text{ }t>0,
\end{equation}

where $\alpha (t)=\cos t/2\sin t,$ $\beta (t)=-1/\sin t,$ and $\gamma
(t)=\cos t/2\sin t.$ It is easy to verify that the convolution operator (\ref%
{ConvOpe}) becomes 
\begin{equation}
U\left( t,\frac{d}{dx}\right) =\sum_{m\geq 0}\frac{1}{m!}\left( i\mu
_{1}(t)\mu _{0}^{{}}\left( t\right) \right) ^{m}\frac{d^{2m}}{dx^{2m}},
\end{equation}

with $\mu _{0}(t)=\sin t$ and $\mu _{1}(t)=\cos t/2.$ It follows from
Corollary 1 that superoscillations hold. See Figure \ref{fig2} (a).
\end{example}

The Green function for the following example was studied by Suslov and
Lanfear in \cite{Lan:Sus}, and in \cite{Aha3} its superoscillations were
studied.

\begin{example}
As explained in \cite{Lan:Sus} the Green function for the Schr\"{o}dinger
equation%
\begin{equation*}
i\frac{\partial \psi }{\partial t}+\frac{1}{4}\frac{\partial ^{2}\psi }{%
\partial x^{2}}\pm tx^{2}\psi =0
\end{equation*}

is of the form 
\begin{equation}
G(x,y,t)=\frac{1}{\sqrt{\pm i\pi \mu _{0}(\pm t)}}\exp \left( \pm i\frac{\mu
_{0}(t)^{\prime }(\pm t)-2xy+\mu _{1}(t)(\pm t)y^{2}}{\mu _{0}(t)(\pm t)}%
\right) ,\text{ }t>0,
\end{equation}

where $\mu _{0}(t)=3^{-2/3}\Gamma \left( \frac{1}{3}\right)
t^{1/2}I_{1/3}\left( \frac{2}{3}t^{3/2}\right) ,$ $\mu _{0}(0)=0,$ $\mu
_{0}^{\prime }(0)=1$ and $\mu _{1}(t)=3^{-1/3}\Gamma \left( \frac{2}{3}%
\right) t^{1/2}I_{-1/3}\left( \frac{2}{3}t^{3/2}\right) ,$ $\mu _{1}^{\prime
}(0)=0,$ $\mu _{1}^{{}}(0)=1,$ and where $I_{v}$ is the modified Besse
function%
\begin{equation}
I_{v}(z)=\left( \frac{z}{2}\right) ^{v}\sum_{k=0}^{\infty }\frac{%
(z^{2}/4)^{k}}{k!\Gamma (v+k+1)}.
\end{equation}

Therefore the Cauchy initial value problem subject to%
\begin{equation*}
\psi (x,0)=e^{ihx}
\end{equation*}%
is given by 
\begin{equation*}
\psi _{h}(x,t)=\frac{1}{\sqrt{2\mu _{1}(t)}}e^{i\left[ \left( \frac{\mu
_{0}^{^{\prime }}(t)}{\mu _{0}(t)}-\frac{1}{\mu _{0}(t)\mu _{1}(t)}\right)
x^{2}+\frac{hx}{\mu _{1}(t)}-\frac{h^{2}\mu _{0}(t)}{4\mu _{1}(t)}\right] }.
\end{equation*}

Also, $\alpha ,$ $\beta $ and $\gamma $ are given by%
\begin{equation*}
\alpha (t)=\frac{\mu _{0}^{\prime }\left( t\right) }{\mu _{0}\left( t\right) 
},\text{ }\beta (t)=-\frac{1}{\mu _{0}\left( t\right) },\text{\ }\gamma (t)=%
\frac{1}{2}\frac{\mu _{1}\left( t\right) }{\mu _{0}\left( t\right) }.
\end{equation*}%
Further, superoscillations hold by Corollary 1.
\end{example}

\begin{example}
The solution for the Cauchy initial value problem for the Caldirola-Kanai
Hamiltonian 
\begin{equation*}
i\frac{\partial \psi }{\partial t}=-\frac{1}{2}e^{-2\lambda t}\frac{\partial
^{2}\psi }{\partial x^{2}}+\frac{1}{2}e^{2\lambda t}x^{2}\psi
\end{equation*}%
satisfying the initial condition 
\begin{equation*}
\psi (x,0)=e^{ihx}
\end{equation*}%
is given by 
\begin{equation*}
\psi _{h}(x,t)=\frac{1}{\sqrt{\mu _{1}(t)}}e^{i\left[ \left( \frac{\mu
_{0}^{^{\prime }}(t)}{\mu _{0}(t)}-\frac{1}{\mu _{0}(t)\mu _{1}(t)}\right)
x^{2}+\frac{hx}{\mu _{1}(t)}-\frac{h^{2}\mu _{0}(t)}{4\mu _{1}(t)}\right] }
\end{equation*}%
where%
\begin{equation*}
\mu _{0}(t)=\frac{\sin (\omega t)}{e^{\lambda t}\omega },\text{ }\mu _{1}(t)=%
\frac{\lambda \sin (\omega t)+\omega \cos (\omega t)}{e^{\lambda t}\omega },%
\text{ }\omega =\sqrt{1-\lambda ^{2}}>0
\end{equation*}

and%
\begin{equation*}
\alpha (t)=\frac{\omega \cos (\omega t)-\lambda \sin (\omega t)}{2\sin
(\omega t)}e^{2\lambda t},\text{ }\beta (t)=-\frac{e^{\lambda t}\omega }{%
\sin (\omega t)},\text{ }\gamma (t)=\frac{\omega \cos (\omega t)+\lambda
\sin (\omega t)}{2\sin (\omega t)}.
\end{equation*}%
Further, superoscillations hold by Corollary 1. See Figure \ref{fig2} (b).
\end{example}

\begin{figure}[h]
\begin{centering}
\subfigure[Approximate real part of the surface.]{\includegraphics[scale=0.2]{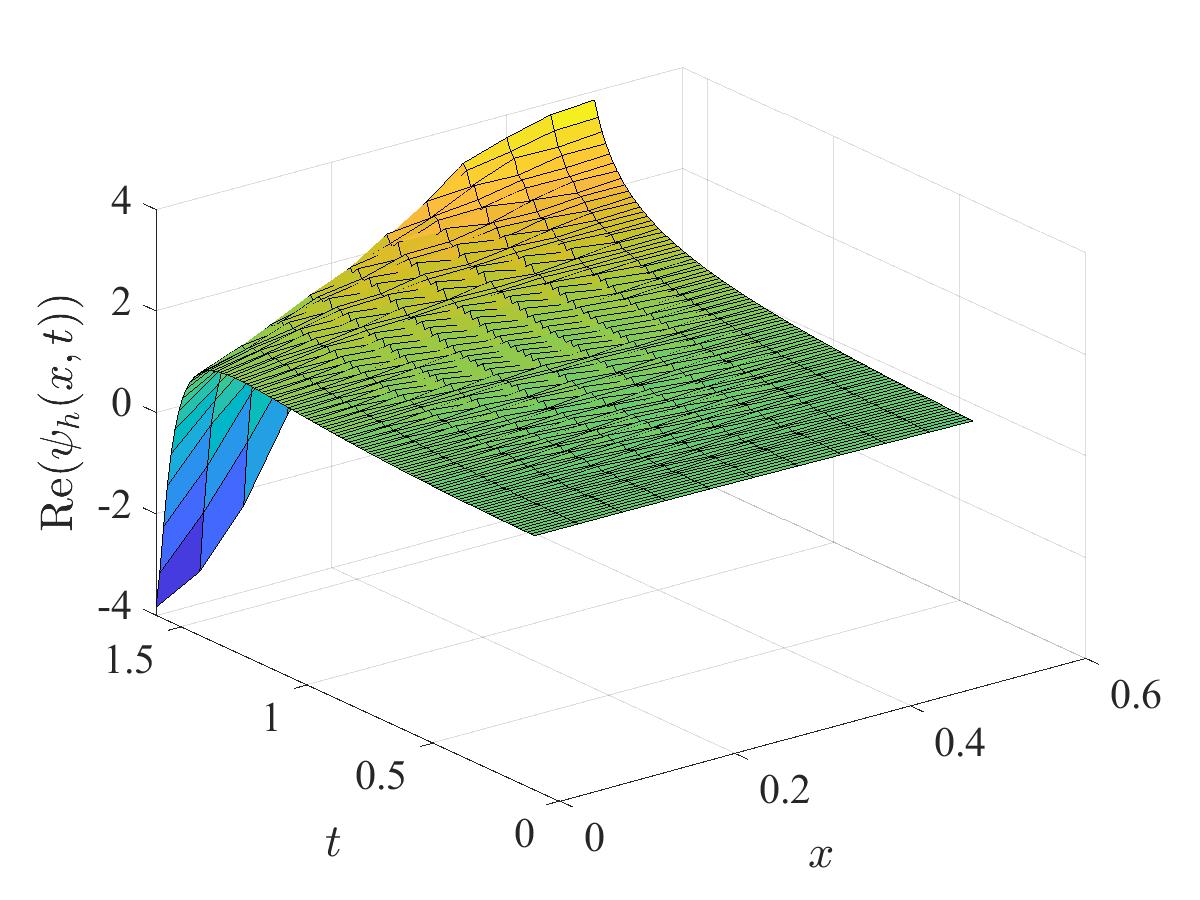}}\subfigure[Exact real part of the surface.]{\includegraphics[scale=0.2]{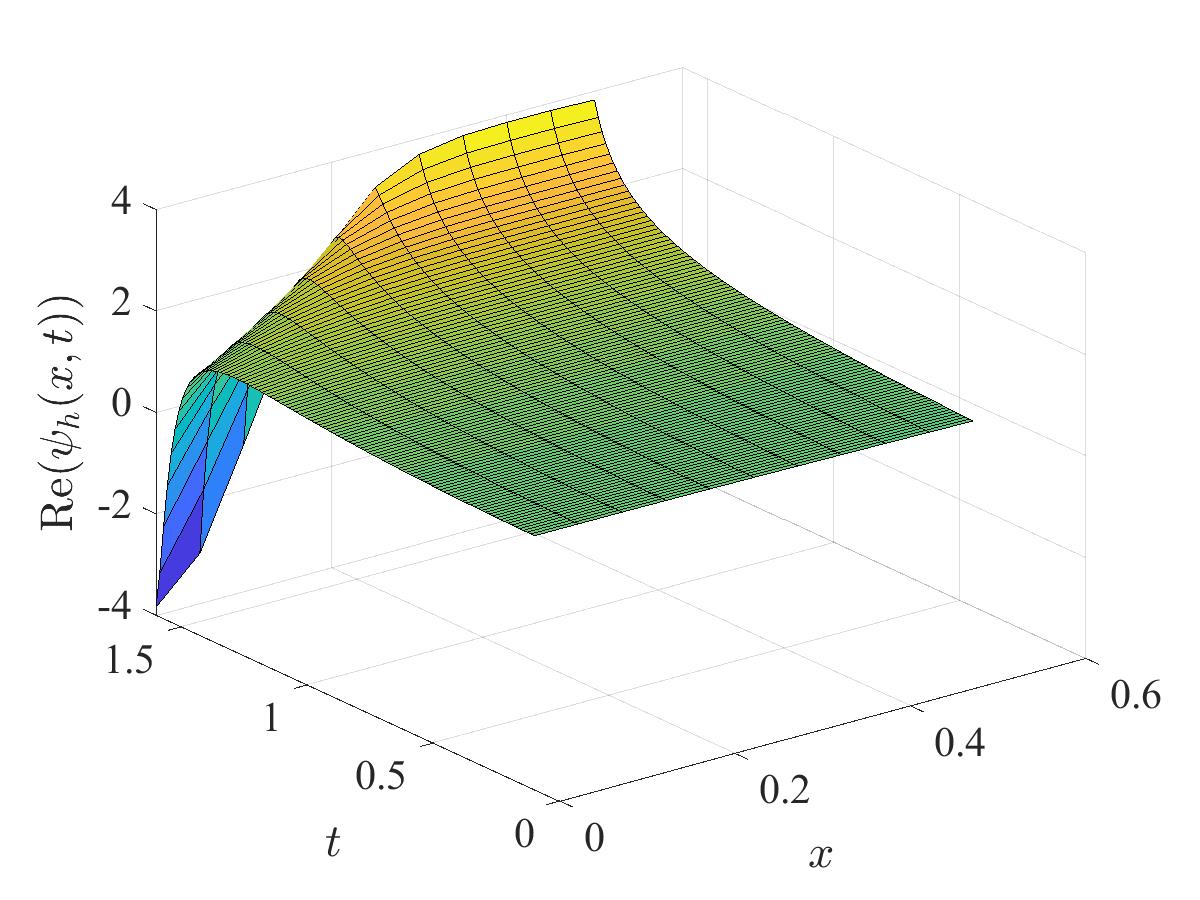}}
\subfigure[Approximate imaginary part of the surface.]{\includegraphics[scale=0.2]{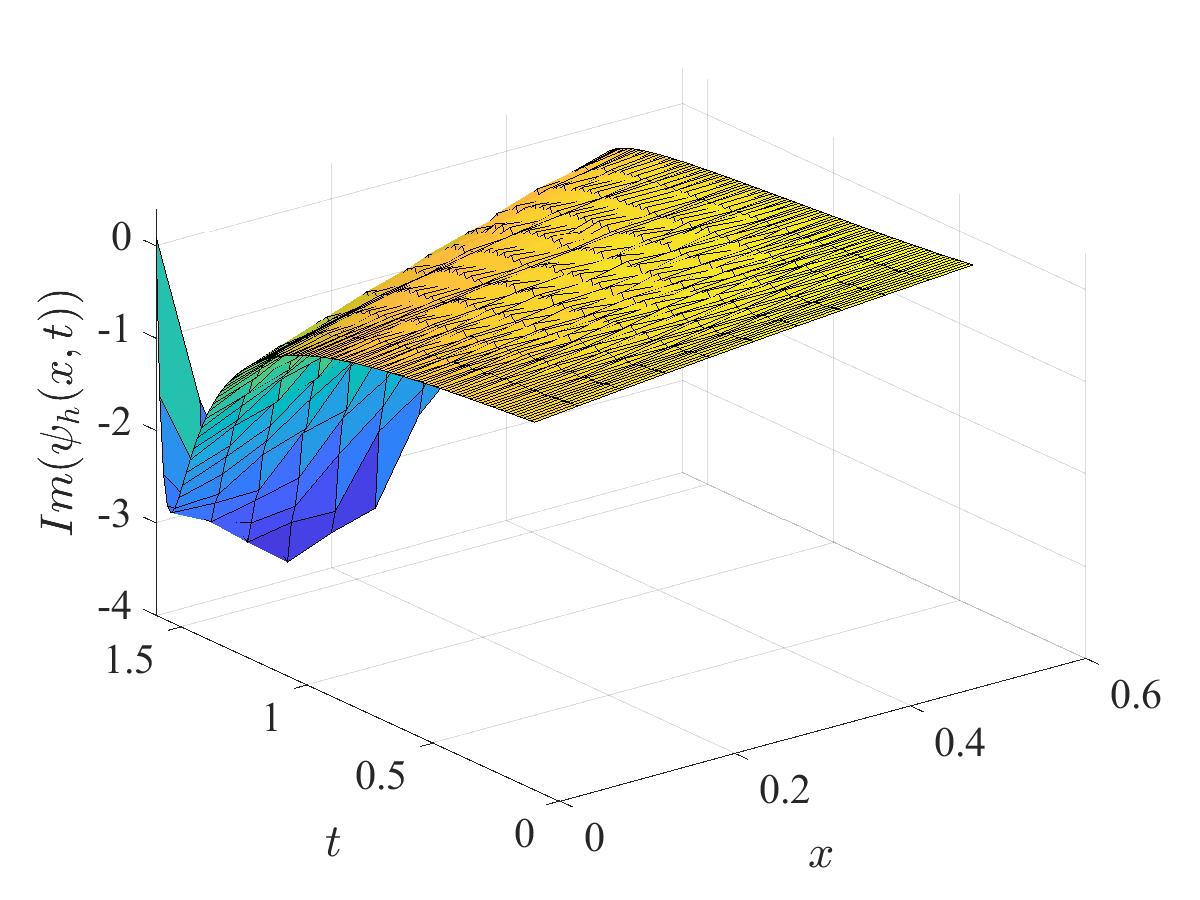}}\subfigure[Exact imaginary part of the surface.]{\includegraphics[scale=0.2]{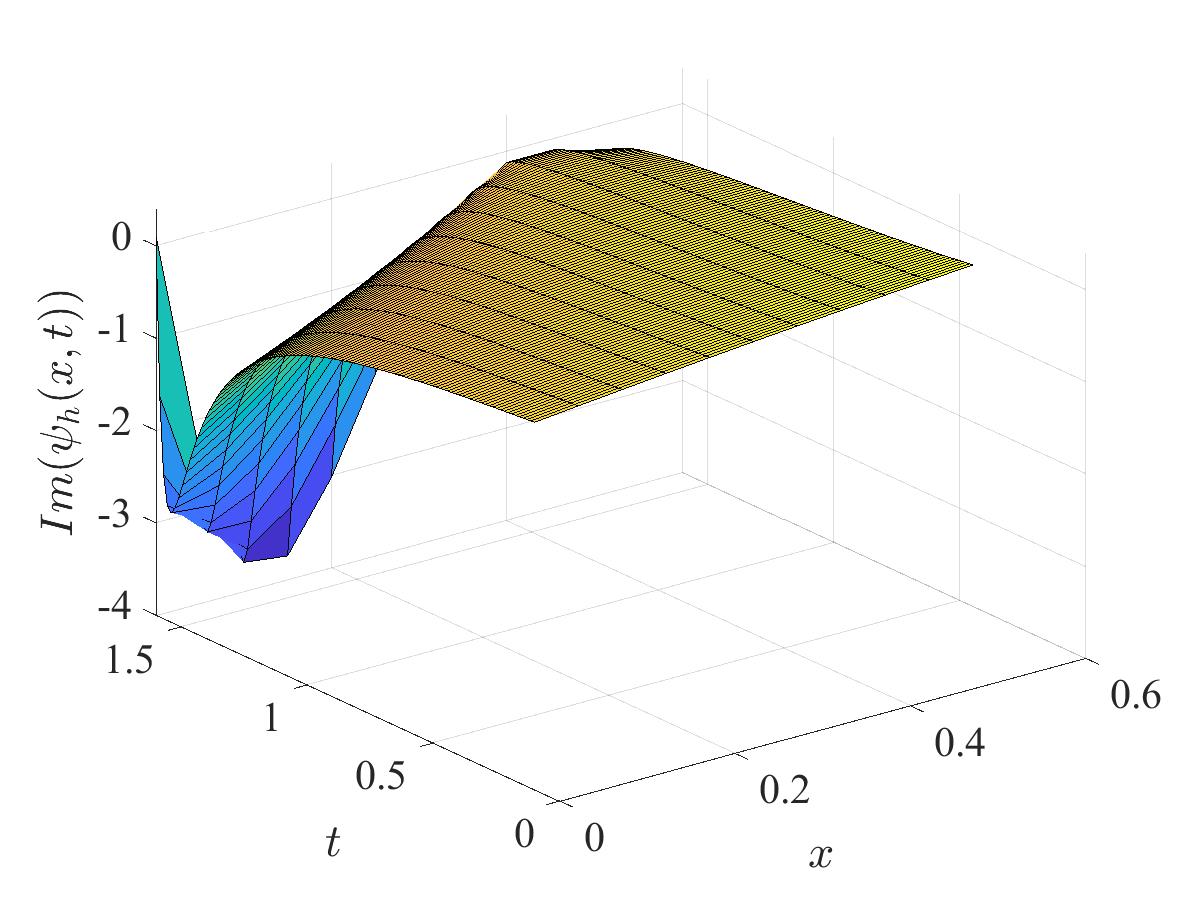}}
\par\end{centering}
\caption{Approximate and exact solutions of the Cauchy initial value problem
for the Caldirola-Kanai Hamiltonian.}
\label{fig3}
\end{figure}

To numerically solve the Cauchy initial value problem for the
Caldirola-Kanai Hamiltonian in Example 3, We used a finite difference over
the space. We also used Runge-Kutta of hybrid order 4 and 5 in MATLAB to
solve the discretized equation over time. The results are shown in Figure %
\ref{fig3}. 

\begin{example}
The solution for the Cauchy initial value problem for the Modified
Caldirola-Kanai Hamiltonian 
\begin{equation*}
i\frac{\partial \psi }{\partial t}=-\frac{\omega _{0}}{2}e^{-2\lambda t}%
\frac{\partial ^{2}\psi }{\partial x^{2}}+\frac{\omega _{0}}{2}e^{2\lambda
t}x^{2}\psi +i\left( 2\lambda x\frac{\partial \psi }{\partial x}+\lambda
\psi \right)
\end{equation*}%
satisfying the initial condition 
\begin{equation*}
\psi (x,0)=e^{ihx}
\end{equation*}%
is given by 
\begin{equation*}
\psi _{h}(x,t)=\sqrt{\frac{\omega e^{\lambda t}}{\omega \cos (\omega
t)-\lambda \sin (\omega t)}}e^{i\left[ \alpha (t)x^{2}-\frac{\left( \beta
(t)x+h)^{2}\right) }{4\gamma (t)}\right] },\text{ }\omega =\sqrt{\omega
_{0}^{2}-\lambda ^{2}}>0,
\end{equation*}%
where%
\begin{equation*}
\mu _{0}(t)=\frac{\omega _{0}\sin (\omega t)}{e^{\lambda t}\omega },
\end{equation*}%
\begin{equation*}
\mu _{1}(t)=\omega \cos (\omega t)-\lambda \sin (\omega t),
\end{equation*}%
\begin{equation*}
\alpha (t)=\frac{\omega \cos (\omega t)-\lambda \sin (\omega t)}{2\omega
_{0}\sin (\omega t)}e^{2\lambda t},\text{ }\beta (t)=-\frac{e^{\lambda
t}\omega }{\omega _{0}\sin (\omega t)}
\end{equation*}%
and%
\begin{equation*}
\gamma (t)=\frac{\omega \cos (\omega t)-\lambda \sin (\omega t)}{2\omega
_{0}\sin (\omega t)}.
\end{equation*}%
Further, superoscillations hold by Corollary 1.
\end{example}

\begin{example}
The solution for the Cauchy initial value problem for the Meiler,
Cordero-Soto, Suslov Hamiltonian 
\begin{equation*}
i\frac{\partial \psi }{\partial t}=-\cos ^{2}(t)\frac{\partial ^{2}\psi }{%
\partial x^{2}}+\sin ^{2}(2t)x^{2}\psi -i\left( \sin (2t)x\frac{\partial
\psi }{\partial x}+\frac{1}{2}\sin (2t)\psi \right)
\end{equation*}%
satisfying the initial condition 
\begin{equation*}
\psi (x,0)=e^{ihx}
\end{equation*}%
is given by%
\begin{equation*}
\psi _{h}(x,t)=\frac{e^{i\left[ \alpha (t)x^{2}-\frac{\left( \beta
(t)x+h)^{2}\right) }{4\gamma (t)}\right] }}{\sqrt{2\cosh (t)\cos (t)+2\sinh
(t)\sin (t)}},\text{ }\omega =\sqrt{\omega _{0}^{2}-\lambda ^{2}}>0,
\end{equation*}

where%
\begin{equation*}
\mu _{0}(t)=\cos (t)\sinh (t)+\cosh (t)\sin (t),
\end{equation*}%
\begin{equation*}
\mu _{1}(t)=\cosh (t)\cos (t)-\sinh (t)\sin (t),
\end{equation*}%
\begin{equation*}
\alpha (t)=\frac{\cosh (t)\cos (t)-\sinh (t)\sin (t)}{2\cos (t)\sinh
(t)+2\cosh (t)\sin (t)},
\end{equation*}%
\begin{equation*}
\beta (t)=-\frac{1}{\cos (t)\sinh (t)+\cosh (t)\sin (t)},
\end{equation*}%
and%
\begin{equation*}
\gamma (t)=\frac{\cosh (t)\cos (t)+\sinh (t)\sin (t)}{\cos (t)\sinh
(t)+\cosh (t)\sin (t)}.
\end{equation*}%
Further, superoscillations hold by Corollary 1.
\end{example}

\begin{example}
The degenerate parametric oscillator of the form 
\begin{eqnarray*}
i\frac{\partial \psi }{\partial t} &=&-\frac{1}{2}\left( 1+\frac{\lambda }{%
\omega }\cos (2\omega t)\right) \frac{\partial ^{2}\psi }{\partial x^{2}}%
+\left( 1-\frac{\lambda }{\omega }\cos (2\omega t)\right) \frac{\omega
^{2}x^{2}}{2}\psi \\
&&-i\lambda \sin (2\omega t)x\frac{\partial \psi }{\partial x}-i\frac{%
\lambda }{2}\sin (2\omega t)\psi
\end{eqnarray*}

satisfying the initial condition 
\begin{equation*}
\psi (x,0)=e^{ihx}
\end{equation*}%
is given by%
\begin{equation*}
\psi _{h}(x,t)=\frac{e^{i\left[ \alpha (t)x^{2}-\frac{\left( \beta
(t)x+h)^{2}\right) }{4\gamma (t)}\right] }}{\sqrt{\sin (\omega t)\cosh
(\lambda t)+\cos (\omega t)\sinh (\lambda t)}},
\end{equation*}

where the characteristic equation is given by the following Ince's type
equation:%
\begin{equation}
\mu ^{\prime \prime }+{\frac{2\lambda \,\omega \,\sin \left( 2\,\omega
\,t\right) }{\omega +\lambda \,\cos \left( 2\,\omega \,t\right) }}\mu
^{\prime }+{\frac{{\omega }^{3}-3\,\omega \,{\lambda }^{2}-\left( {\omega }%
^{2}\lambda +{\lambda }^{3}\right) \cos \left( 2\,\omega \,t\right) }{\omega
+\lambda \,\cos \left( 2\,\omega \,t\right) }}\mu =0.  \label{inceeq}
\end{equation}

Two independent solutions for (\ref{inceeq})\ are given by (see appendix for
details)%
\begin{equation*}
\mu _{0}(t)=\sin (\omega t)\cosh (\lambda t)+\cos (\omega t)\sinh (\lambda
t),
\end{equation*}%
\begin{equation*}
\mu _{1}(t)=\sin (\omega t)\sinh (\lambda t)+\cos (\omega t)\cosh (\lambda
t),
\end{equation*}%
\begin{equation*}
\alpha (t)=\frac{\omega (\sinh (\lambda t)\sin (\omega t)-\cosh (\lambda
t)\cos (\omega t))}{2(\sin (\omega t)\cosh (\lambda t)+\cos (\omega t)\sinh
(\lambda t))},
\end{equation*}%
\begin{equation*}
\beta (t)=-\frac{\omega }{\sin (\omega t)\cosh (\lambda t)+\cos (\omega
t)\sinh (\lambda t)},
\end{equation*}%
and%
\begin{equation*}
\gamma (t)=-\frac{\omega (\sinh (\lambda t)\sin (\omega t)+\cosh (\lambda
t)\cos (\omega t))}{2(\sin (\omega t)\cosh (\lambda t)+\cos (\omega t)\sinh
(\lambda t))}.
\end{equation*}%
\ \ \ \ \ \ \ \ \ \ \ \ \ \ \ \ Further, superoscillations hold by Corollary
1.
\end{example}

\begin{example}
The quantum harmonic oscillator 
\begin{equation*}
i\frac{\partial \psi }{\partial t}=-a(t)\frac{\partial ^{2}\psi }{\partial
x^{2}}+\frac{b(t)}{2}x^{2}\psi ,
\end{equation*}

with%
\begin{equation*}
a(t)=\frac{\Omega ^{2}\cos (\Omega t)-\gamma \sin (\Omega t)\tan (\Omega t)}{%
\cosh (\gamma t)(\cos (\gamma t)\cosh (\gamma t)-2\gamma )},\text{ }b(t)=-%
\frac{\omega ^{2}}{4a(t)},
\end{equation*}

and $\Omega =\sqrt{\omega ^{2}-\gamma ^{2}},$ its Green function is given by 
\begin{equation}
G(x,y,t)=\sqrt{\frac{m_{0}\Omega \cosh (\gamma t)}{2\pi i\sin (\Omega t)}}%
\exp \left( i\alpha (t)x^{2}+\beta (t)xy+\gamma (t)y^{2}\right) ,\text{ }t>0,
\end{equation}

where $\alpha (t)=(\cosh (\gamma t)(m_{0}\Omega \cosh (\gamma t)\cos (\Omega
t)-\gamma ))/2\sin (\Omega t),$ $\beta (t)=-m_{0}\Omega \cosh (\gamma
t)/2\pi \sin (\Omega t),$ and $\gamma (t)=-m_{0}\Omega \cos (\gamma t)/2\sin
(\Omega t).$

It follows from the Corollary 1 that superoscillations hold.
\end{example}

\section{Other type of superoscillating data}

\begin{corollary}
(See \cite{Aha3}) Let $h>1,$ $p$ even, and let $L$ be a real positive
number. Then, for all $x\in \lbrack -L,L],$ the sequence%
\begin{equation*}
Y_{n}(x)=\sum_{k=0}^{n}C_{k}(n,h)e^{ix\left( -i(1-2k/n)\right) ^{p}}
\end{equation*}

is $e^{ix(-h)^{p}}$-superoscilating, i.e. we have%
\begin{equation*}
\lim_{n\rightarrow \infty }Y_{n}(x)=e^{ix(-h)^{p}}.
\end{equation*}
\end{corollary}

\begin{theorem}
Let $p=2r$ even. Consider the superoscillating function%
\begin{equation*}
Y_{n}(x)=\sum_{k=0}^{n}C_{k}(n,h)e^{ix\left( -i(1-2k/n)^{p}\right) }
\end{equation*}%
Then, the solution of the Cauchy initial value problem satisfying the
initial condition 
\begin{eqnarray}
i\frac{\partial \psi }{\partial t} &=&-a\left( t\right) \partial
_{x}^{2}\psi +b\left( t\right) x^{2}\psi -ic\left( t\right) x\partial
_{x}\psi -id\left( t\right) \psi ,\quad  \label{SEOscillations} \\
\psi (x,0) &=&Y_{n}(x),  \label{DataSEOscillations}
\end{eqnarray}

is given by 
\begin{equation}
\psi _{n}(x,t)=\sum_{k=0}^{n}\frac{C_{k}(n,h)e^{i(\alpha (t)x^{2}+\delta
(t)x+\kappa (t))}e^{-i[\beta (t)x+\varepsilon (t)+(-i(1-2k/n)^{p})]^{2}}}{%
\sqrt{2\mu _{0}(t)\gamma (t)}}, Im(\gamma (t))\leq 0.  \label{phin}
\end{equation}

Moreover, if we set $\psi (x,t)=\lim_{n\rightarrow \infty }\psi _{n}(x,t),$
then%
\begin{equation}
\psi (x,t)=\frac{e^{i(4\alpha -\beta ^{2})x^{2}/4\gamma }}{\sqrt{2\gamma \mu 
}}U\left( t,\frac{d}{dx}\right) e^{i\beta (-h)^{2r}x/2\gamma },
\end{equation}

where 
\begin{equation}
U\left( t,\frac{d}{dx}\right) :=\sum_{m\geq 0}^{{}}\frac{1}{m!}\left[ \frac{%
i\gamma }{\beta ^{2}\left( -\left( 1-2k/n\right) ^{2}\right) ^{2r}}\right]
^{m}\frac{d^{2m}}{dx^{2m}}.  \label{operatorUY}
\end{equation}
\end{theorem}

Proof. We have assumed that $Im(\gamma (t))\leq 0.$The solution for (\ref%
{SEOscillations})- (\ref{DataSEOscillations}) by Lemma 1 is given by%
\begin{eqnarray*}
\phi _{1-\frac{2k}{n}}(x,t) &=&\frac{1}{\sqrt{2\pi i\mu _{0}}}%
\int_{{}}^{{}}e^{i(\alpha x^{2}+\beta xy+\gamma y^{2})+iy(-i(1-2k/n))^{2r}}dy
\\
&=&\frac{1}{\sqrt{2\gamma \mu _{0}}}e^{i(4\alpha -\beta ^{2})x^{2}/4\gamma
(t)}e^{-i\beta (-\left( 1-2k/n\right) ^{2})^{r}x/2\gamma (t)}e^{i(-\left(
1-2k/n\right) ^{2})^{2r}/4\gamma (t)}.
\end{eqnarray*}

Similarly we can find that

\begin{equation*}
\phi _{h}(x,t)=\frac{1}{\sqrt{2\gamma \mu _{0}}}e^{i(4\alpha -\beta
^{2})x^{2}/4\gamma (t)}e^{-i\beta (-h^{2})^{r}x/2\gamma
(t)}e^{-i(-h^{2})^{2r}/4\gamma (t)}.
\end{equation*}

We proceed as in previous sections and we can rewrite the latter term as%
\begin{equation*}
\phi _{h}(x,t)=\frac{1}{\sqrt{2\pi i\mu _{0}(t)}}e^{i(4\alpha -\beta
^{2})x^{2}/4\gamma (t)}\sum_{m\geq 0}^{{}}\frac{1}{m!}\left( \frac{-i}{%
4\gamma }\right) ^{m}\left( h^{2r}\right) ^{2m}e^{-i\beta
(-h^{2})^{r}x/2\gamma (t)}.
\end{equation*}

Also, it is easy to prove that the following expression holds%
\begin{equation*}
\left( h^{2r}\right) ^{2m}e^{-i\beta (-h^{2})^{r}x/2\gamma (t)}=\left[ \frac{%
2\gamma }{-i\beta (-h^{2})^{r}}\right] ^{2m}\frac{d^{2m}}{dx^{2m}}e^{-i\beta
(-h^{2})^{r}x/2\gamma (t)}.
\end{equation*}

Therefore, we obtain

\begin{equation*}
\phi _{h}(x,t)=\frac{e^{i(4\alpha -\beta ^{2})x^{2}/4\gamma (t)}}{\sqrt{%
2\gamma \mu _{0}(t)}}\sum_{m\geq 0}^{{}}\frac{1}{m!}\left[ \frac{i\gamma }{%
\beta ^{2}\left( -h^{2}\right) ^{2r}}\right] ^{m}\frac{d^{2m}}{dx^{2m}}%
e^{-i\beta (-h^{2})^{r}x/2\gamma (t)}.
\end{equation*}

From the latter we define $U\left( t,\frac{d}{dx}\right) $ as (\ref%
{operatorUY}), so we also obtain

\begin{equation*}
\phi _{1-\frac{2k}{n}}(x,t)=\frac{e^{i(4\alpha -\beta ^{2})x^{2}/4\gamma (t)}%
}{\sqrt{2\gamma \mu _{0}(t)}}U\left( t,\frac{d}{dx}\right) e^{-i\beta
(-\left( 1-2k/n\right) ^{2})^{r}x/2\gamma (t)}.
\end{equation*}

Therefore, the solution $\psi _{n}(x,t)$ for (\ref{SEOscillations})-(\ref%
{DataSEOscillations}) by the superposition principle is given by 
\begin{eqnarray*}
\psi _{n}(x,t) &=&\sum_{k=0}^{n}C_{k}(n,h)\phi _{1-\frac{2k}{n}}(x,t) \\
&=&\sum_{k=0}^{n}C_{k}(n,h)\frac{e^{i(4\alpha -\beta ^{2})x^{2}/4\gamma (t)}%
}{\sqrt{2\gamma \mu _{0}(t)}}U\left( t,\frac{d}{dx}\right) e^{-i\beta
(-\left( 1-2k/n\right) ^{2})^{r}x/2\gamma (t)}.
\end{eqnarray*}

Further, taking the limit and using Lemma 2 by Aharonov et al. and by
Corollary 2 we would obtain%
\begin{eqnarray*}
\lim_{n\rightarrow \infty }\psi _{n}(x,t) &=&\lim_{n\rightarrow \infty
}\sum_{k=0}^{n}C_{k}(n,h)\frac{e^{i(4\alpha -\beta ^{2})x^{2}/4\gamma }}{%
\sqrt{2\gamma \mu _{0}}}U\left( t,\frac{d}{dx}\right) e^{-i\beta (-i\left(
1-2k/n\right) )^{2r}x/2\gamma } \\
&=&\frac{e^{i(4\alpha -\beta ^{2})x^{2}/4\gamma }}{\sqrt{2\gamma \mu _{0}}}%
U\left( t,\frac{d}{dx}\right) e^{-i\beta (-h)^{2r}x/2\gamma }.
\end{eqnarray*}

This finishes the proof.

A similar result holds for $p=2r+1,$ $r\in 
\mathbb{Z}
$ and the sequence is $Z_{n}=\sum_{k=o}^{n}C_{k}(n,a)e^{x(-i(1-2k/n))^{p}}.$

\begin{theorem}
Let $p=2r+1$ be odd. Consider the superoscillating function%
\begin{equation*}
Z_{n}(x)=\sum_{k=0}^{n}C_{k}(n,h)e^{x\left( -i(1-2k/n)^{p}\right) }.
\end{equation*}%
Then, the solution of the Cauchy initial value problem satisfying the
initial condition 
\begin{eqnarray}
i\frac{\partial \psi }{\partial t} &=&-a\left( t\right) \partial
_{x}^{2}\psi +b\left( t\right) x^{2}\psi -ic\left( t\right) x\partial
_{x}\psi -id\left( t\right) \psi ,\quad  \label{SEOscillations} \\
\psi (x,0) &=&Z_{n}(x)  \label{DataSEOscillations}
\end{eqnarray}

is given by 
\begin{equation}
\psi _{n}(x,t)=\sum_{k=0}^{n}\frac{C_{k}(n,h)e^{i(4\alpha -\beta
^{2})x^{2}/4\gamma }}{\sqrt{2\mu _{0}(t)\gamma (t)}}U\left( t,\frac{d}{dx}%
\right) e^{-i\beta (-ih)^{2r+1}x/2\gamma },Im(\gamma (t))\leq 0.
\label{phin}
\end{equation}

Moreover, if we set $\psi (x,t)=\lim_{n\rightarrow \infty }\psi _{n}(x,t),$
then%
\begin{equation}
\psi (x,t)=\frac{e^{i(4\alpha -\beta ^{2})x^{2}/4\gamma }}{\sqrt{2\gamma \mu
_{0}}}U\left( t,\frac{d}{dx}\right) e^{-i\beta (-ih)^{2r+1}x/2\gamma },
\end{equation}

where 
\begin{equation}
U\left( t,\frac{d}{dx}\right) :=\sum_{m\geq 0}^{{}}\frac{1}{m!}\left[ \frac{%
ih^{2r}\gamma }{-\beta ^{2}(-h^{2})^{2r+1}}\right] ^{m}\frac{d^{2m}}{dx^{2m}}%
.
\end{equation}
\end{theorem}

\begin{conclusion}
For less than a century, the study of superoscillations in physical systems
has proven to be a most puzzling and exciting phenomenon. Originally as a
natural consequence of the principles of Fourier analysis, globally
band-limited signals (e.g electrical, audio, etc) do not convey information
beyond that of the smallest period of their Fourier components; as a result,
it was thought that weakened measurement interactions that did not disturb
the system produced no data. However \cite{Tollakasen} has shown, to the
contrary, that this is not the case. In a study \cite{Aha0}, \cite{Aha01},
Aharonov and collaborators showed that these weak valued measurement
interactions resulted in weak values that lead in a new physical effect
termed superoscillations. In particular, the waveforms that characterize
these superoscillations are currently under consideration in many
engineering applications such as the theory of super-resolution in optics.
Due to growth of study in these areas and their applications, we encourage
and defer the reader to the work of Berry et al, \cite{Berry1}-\cite{Berry7}
and \cite{Lindberg} also contains an excellent survey of the most recent
applications in the areas of engineering and technology.

In this work, we have studied the persistence in time of superoscillations
for the Schr\"{o}dinger equation of the form (\ref{SE1}); this is probably
the most general time dependent quadratic Hamiltonian for which
superoscillations has been proven. In order to prove the persistence of
superoscillations we have defined explicitly a pseudodifferential operator
in terms of solutions of a Riccati system associated with the variable
coefficients of the Hamiltonian. We have also solved explicitly the Cauchy
initial value problem with oscillatory initial data in terms of a Riccati
system. The pseudodifferential operator is defined on a space of entire
functions. Particular examples include Caldirola-Kanai, modified
Caldirola-Kanai, degenerate parametric harmonic oscillator and Meiler,
Cordero-Soto, Suslov Hamiltonians.
\end{conclusion}

\begin{acknowledgement}
This research is currently supported by NSF DMS\#1620196, NSF DMS\#1620268.
It was partially funded by the program of the Mathematical Association of
America funded by the NSF Grant DMS-1652506 and College of Sciences Research
Enhancement Seed Grants Program at UTRGV. One of the authors (E.S.) is
supported by the Simons Foundation \#316295. On behalf of all authors, the
corresponding author states that there is no conflict of interest.
\end{acknowledgement}

\section{Appendix A: Solutions for Ince's Type Equation (\protect\ref{inceeq}%
)}

In this appendix we review how to solve Ince's equation \eqref{inceeq} using
the Hamiltonian Algebrization procedure and the Kovacic Algorithm, see \cite%
{Acosta-Suazo} for more details. By properties of double angle, we can write
the equation \eqref{inceeq} in terms of $\tan (\omega t)$. For instance, we
can consider as its differential field $K=\mathbb{C}(\tan \omega t)$. After
the Hamiltonian change of variable $\tau =\tan \omega t$ we obtain $\alpha
=\omega ^{2}(1+\tau ^{2})^{2},$ and by the Hamiltonian Algebrization
procedure we get the algebraic form of \eqref{inceeq} as follows 
\begin{equation*}
\begin{array}{l}
\partial _{\tau }^{2}\widehat{\mu }+\varphi _{1}(\tau )\partial _{\tau }%
\widehat{\mu }+\varphi _{0}(\tau )\widehat{\mu }=0,\quad \varphi _{1}(\tau )=%
{\frac{2(\lambda -\omega )\tau ^{3}-(3\lambda +\omega )\tau }{(1+\tau
^{2})\left( (\lambda -\omega )\tau ^{2}-\lambda -\omega \right) }}, \\ 
\\ 
\varphi _{0}(\tau )=-{\frac{\left( {\omega }^{3}-3\,\omega \,{\lambda }^{2}+{%
\omega }^{2}\lambda +{\lambda }^{3}\right) {\tau }^{2}+{\omega }%
^{3}-3\,\omega \,{\ \lambda }^{2}-{\omega }^{2}\lambda -{\lambda }^{3}}{%
\left( 1+{\tau }^{2}\right) ^{2}\left( \left( \lambda -\omega \right) {\tau }%
^{2}-\lambda -\omega \right) {\omega }^{2}}}.%
\end{array}%
\end{equation*}%
We can eliminate one parameter through the change $\lambda =\kappa \omega $;
thus, our algebraic form becomes 
\begin{equation}
\begin{array}{l}
\partial _{\tau }^{2}\widehat{\mu }+\varphi _{1}(\tau )\partial _{\tau }%
\widehat{\mu }+\varphi _{0}(\tau )\widehat{\mu }=0,\quad \varphi _{1}(\tau )=%
{\frac{2(\kappa -1)\tau ^{3}-(3\kappa +1)\tau }{(1+\tau ^{2})\left( (\kappa
-1)\tau ^{2}-\kappa -1\right) }}, \\ 
\\ 
\varphi _{0}(\tau )=-{\frac{(1-3\kappa ^{2}+\kappa +\kappa ^{3})\tau
^{2}+1-3\kappa ^{2}-\kappa -\kappa ^{3}}{(1+\tau ^{2})^{2}\left( (\kappa
-1)\tau ^{2}-\kappa -1\right) }},\quad \kappa \neq 1.%
\end{array}
\label{algformincgen}
\end{equation}%
We can transform the equation \eqref{algformincgen} into

\begin{equation}
\begin{array}{l}
\partial _{\tau }^{2}y=ry,\quad \widehat{\mu }(\tau )=y{\frac{\sqrt{(\kappa
-1){\tau }^{2}-1-\kappa }}{{1+{\tau }^{2}}}} \\ 
\\ 
r={\frac{\left( \left( -4{\kappa }^{3}-4\kappa +7\kappa ^{2}+{\kappa }%
^{4}\right) {\tau }^{4}+\left( 10\kappa ^{2}-2{\kappa }^{4}\right) {\ \tau }%
^{2}+4\kappa +7\kappa ^{2}+4{\kappa }^{3}+{\kappa }^{4}\right) }{\left( 1+{%
\tau }^{2}\right) ^{2}\left( \left( -1+\kappa \right) {\tau }^{2}-1-\kappa
\right) ^{2}}}.%
\end{array}
\label{alformincred}
\end{equation}%
We see that the poles of $r$ are given by the set $\Gamma =\left\{ i,-i,%
\sqrt{\frac{\kappa +1}{\kappa -1}},-\sqrt{\frac{\kappa +1}{\kappa -1}}%
,\infty \right\} $, $\circ r_{c}=2,\forall c\in \Gamma $, which implies that
equation \eqref{alformincred} could be solved using one of the cases 1, 2, 3
or 4 of the Kovacic's algorithm. We discard case one (see \cite{Acosta-Suazo}
for details), and by step two and step three of the Kovacic's algorithm we
obtain the general solution of \eqref{alformincred}: 
\begin{equation}
y=C_{1}e^{-\kappa \arctan \tau }(\tau -1)\sqrt{1+\tau ^{2}}+C_{2}e^{\kappa
\arctan \tau }(\tau +1)\sqrt{1+\tau ^{2}},  \label{solgenince}
\end{equation}%
for instance $\mathrm{DGal}(\widehat{L}/\widehat{K})=\mathbb{D}_{\infty }$,
that is, the infinite dihedral group for any $\kappa \neq 0$. Now, the
general solution for equation \eqref{algformincgen} is given by 
\begin{equation}
\widehat{\mu }(\tau )=C_{1}{\frac{e^{-\kappa \arctan \tau }(\tau -1)}{\sqrt{%
1+\tau ^{2}}}}+C_{2}{\frac{e^{\kappa \arctan \tau }(\tau +1)}{\sqrt{1+\tau
^{2}}}};  \label{solgenince2}
\end{equation}%
for instance the differential Galois group for the algebrized characteristic
equation \eqref{algformincgen} is also the dihedral infinite group $\mathbb{D%
}_{\infty }$ for any value of $\kappa \neq 0$. Recalling that $\tau =\tan
\lambda t$ and $\lambda =\kappa \omega $, we get the general solution of the
characteristic equation 
\begin{equation*}
\mu (t)=C_{1}e^{-\lambda t}(\sin \omega t-\cos \omega t)+C_{2}e^{\lambda
t}(\sin \omega t+\cos \omega t),
\end{equation*}%
which can also be written as 
\begin{equation*}
\mu (t)=(C_{1}+C_{2})(\sinh \lambda t\cos \omega t+\cosh \lambda t\sin
\omega t
\end{equation*}%
\begin{equation*}
+(C_{2}-C_{1})\sinh \lambda t\sin \omega t+\cosh \lambda t\cos \omega t,
\end{equation*}%
and its differential Galois group is also the dihedral infinite group, i.e., 
$\mathrm{DGal}(L/K)=\mathbb{D}_{\infty }$. Now, we find $\mu _{0}(t)$ and $%
\mu _{1}(t)$ satisfying the initial conditions (\ref{inceeq}), \ obtaining 
\begin{equation*}
\mu _{0}(t)=\sinh \lambda t\cos \omega t+\cosh \lambda t\sin \omega t,\quad
C_{1}=C_{2}=\frac{1}{2},
\end{equation*}%
\begin{equation*}
\mu _{1}(t)=\sinh \lambda t\sin \omega t+\cosh \lambda t\cos \omega t,\quad
-C_{1}=C_{2}=\frac{1}{2}.
\end{equation*}

\end{document}